\documentclass[a4paper,11pt]{article}
\pdfoutput=1
\newcommand{\orcid}[1]{\href{https://orcid.org/#1}{\includegraphics[scale=.6]{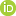}}}
\usepackage{graphicx,epsfig}
\usepackage{grffile}
\usepackage{amsmath,amssymb}
\usepackage{array}
\usepackage[utf8]{inputenc}
\usepackage[usenames, dvipsnames]{color}
\usepackage{slashed}
\usepackage[normalem]{ulem}
\usepackage{jheppub}
\usepackage{url}
\usepackage[utf8]{inputenc} 
\usepackage{mathtools}
\usepackage{lineno}

\DeclarePairedDelimiterX\braket[2]{\langle}{\rangle}{#1 \delimsize\vert #2}
\usepackage{enumerate}

\usepackage{float}
\usepackage[caption = false]{subfig}

\numberwithin{equation}{section}

\usepackage{xcolor}

\title{\Large\bf\boldmath
Quantum decoherence and relaxation in long-baseline neutrino data
}
\date{\today}
\author[a]{A.~L.~G.~Gomes,}
\author[a]{R.~A.~Gomes\,\orcid{0000-0003-0278-4876},}
\author[b]{and O.~L.~G.~Peres\,\orcid{0000-0003-2104-8460}\,}
\affiliation[a]{Instituto de F\'isica, Universidade Federal de Goi\'as, 74690-900, Goi\^ania, GO, Brazil}
\affiliation[b]{Instituto de F\'isica Gleb Wataghin, UNICAMP, 13083-859, Campinas, SP, Brazil}

\emailAdd{abnergomes@ufg.br}
\emailAdd{ragomes@ufg.br}
\emailAdd{orlandop@unicamp.br}

\abstract{
We investigate the effect of quantum decoherence and relaxation in neutrino oscillations using MINOS and T2K data. The formalism of open quantum systems is used to describe the interaction of a neutrino system with the environment, where the strength of the interaction is regulated by a decoherence parameter $\Gamma$. We assume an energy dependence parameterized by $\Gamma = \gamma_0 (E/\mbox{GeV})^n$, with $n=-2,0,+2$, and consider three different scenarios, allowing the investigation of the effect of relaxation and of constraining the solar and atmospheric sectors to the same decoherence parameter. The MINOS and T2K data present a complementary behavior, with regard to our theoretical model, resulting in a better sensitivity for $n = +2$ and $n = -2$, respectively. We perform a combined analyses of both experimental data, which also include a reactor constraint on $\sin^2 \theta_{13}$, and observe an independence of the results to the scenarios we investigate. As highlight of our analyses we obtain the best limit on $\gamma_0$ for the energy dependence of $n = -2$, reporting an upper bound of $1.7 \times 10^{-23}$~GeV, at the 90\% confidence level.}

\begin{document}

\maketitle
\flushbottom

\section{Introduction}

The discovery of neutrino oscillation~\cite{nobelprize1} about 20 years ago and consequently the fact that neutrinos are massive particles opened a window to new investigations in neutrino physics. The neutrino oscillation phenomenon arises from a quantum effect of interference among different neutrino mass eigenstates~\cite{GonzalezGarcia:2002dz}. An interesting possibility of investigation is the neutrino quantum decoherence and relaxation, which can affect the interference in oscillating systems~\cite{Ellis:1983jz}.
Quantum decoherence and relaxation, in general, could be originated by: (i) an intrinsic way, when we have a broadening of the width of the wave packet, and (ii) an extrinsic way, when we have an interaction of the neutrino system with the environment, inducing changes in the neutrino evolution. The investigation of the second type is the goal of this work, which can be described by the known Lindblad equation or, being historically correct~\cite{2017OSID...2440001C}, the Gorini-Kossakowski-Sudarshan-Lindblad (GSKL) master equation~\cite{Lindblad1976, Gorini:1975nb}. This equation has been applied to particle physics for a longtime and more recently to neutrino physics.

In this picture many new parameters arise from the neutrino evolution, opening several possibilities to investigate the decoherence and relaxation. Under the neutrino oscillation framework, the decoherence (relaxation) parameters affect the oscillatory (non-oscillatory) terms of the probability~\cite{Oliveira:2014jsa}.
The general scenario of decoherence and relaxation is known as dissipation effect which behavior is similar to the neutrino decay scenario~\cite{Gomes:2014yua}. 
Previous investigations, considering this assumption as a starting point, have constrained decoherence models, using atmospheric neutrinos~\cite{Lisi:2000zt,Coloma:2018idr,Ahlers2018}, accelerator neutrinos~\cite{Gago:2000nv,Gago:2000qc,Oliveira:2010zzd, deOliveira:2013dia,
Oliveira:2014jsa,Oliveira:2016asf,Gomes:2016ixi,Coelho:2017byq,Coelho:2017zes,Carpio:2017nui,Gomes:2018inp}, and solar/reactor neutrinos~\cite{Gago:2002na,Fogli:2007tx,Gomes:2016ixi,deHolanda:2019tuf,deGouvea:2020hfl,deGouvea:2021uvg,JUNO:2021ydg,DeRomeri:2023dht,DEsposito:2023psn}. Recently, the decoherence was also proposed to explain the LSND anomaly~\cite{Barenboim:2004wu,Farzan:2008zv,Bakhti:2015dca,Dixit:2018gjc} and a possible incompatibility in the experimental measurement of the mixing angle $\theta_{23}$~\cite{Coelho:2017zes} among NO$\nu$A~\cite{NOvA:2018gge} and T2K~\cite{Abe:2017vif}.  The kind of decoherence we are interested in this analysis could arise, for instance, from quantum gravity effects~\cite{Lisi:2000zt,Benatti:2001fa,Barenboim:2004wu,Mavromatos:2006yn,Sakharov:2009rn,DEsposito:2023psn}. Motivated by this hypothesis, we can parameterize the decoherence with an energy dependence given by a power-law~\cite{Gago:2000qc,Lisi:2000zt,Farzan:2008zv,Oliveira:2013nua,Bakhti:2015dca,Gomes:2018inp,Coloma:2018idr,Lambiase:2023pxd}.

The precision measurement of $\theta_{13}$ by reactor neutrino experiments~\cite{Adey:2018zwh,Abe:2011fz} allows the investigation of CP violation in the leptonic sector, as well as the neutrino mass ordering. It also allows studies about the possible effect of the decoherence and relaxation on the unanswered issues in neutrino oscillation, as shown by Ref.~\cite{Carpio:2018gum}. Other possibilities, such as CPT violation due to quantum decoherence, are also discussed in Refs.~\cite{Barenboim:2004wu,Capolupo:2018hrp,Carrasco:2018sca,Buoninfante:2020iyr}.

In order to contribute to this active field of investigation, we aim to present new constraints to the decoherence and relaxation. One of our goals in this study is to discuss the decoherence and relaxation effects under the oscillation parameters. For that purpose, we assume a framework of three-flavors neutrino oscillation obeying the normal mass ordering.

This article is organized as follows. In Section~\ref{sec:theo} we introduce the theoretical development of the neutrino oscillation described by the Lindblad dynamics. We also present the proposed scenarios (Section~\ref{our-scenario}) and discuss the effect of the energy dependence on the decoherence and relaxation parameters in the oscillation probability. Next, in Section~\ref{sec:analysis}, we present the $\chi^2$ analyses developed for MINOS and T2K dataset. 
In Section~\ref{sec:results} we first show the results of our analysis for MINOS, T2K, and their combination, considering each scenario investigated, and the effect of the inclusion of a reactor constraint. We then compare our results with the bounds previously reported in the literature.
Finally, we summarize this study and give our conclusions in Section~\ref{sec:conclusions}. The Appendix~\ref{deco-prop} introduces some important properties of the neutrino system in the light of the Lindblad dynamics and a detailed description of the computation of the probability function. The Appendix~\ref{app:validation} describes the validation method to obtain the allowed regions for the parameters of the standard oscillation scenario.

\section{Phenomenological model and scenarios}
\label{sec:theo}

The description of open quantum systems has the foundations of any non-strong interacting environment and with Markovian behavior. That formalism can be described by the Gorini-Kossakowski-Sudarshan-Lindblad equation~\cite{Gorini:1975nb,Lindblad1976}. In that equation, the environmental influence can be described as decoherence and relaxation effects. In the Lindblad formalism all effects are described by matrix density format. Our approach will be {\it phenomenological} for the Lindblad operator, which should have complete positivity, but otherwise can contain any form for the elements of the operator. As we will describe later, in more detail, we will examine  some scenarios, allowing relaxation and/or decoherence effects. 

\subsection{Open quantum system formalism for neutrinos}

The neutrino phenomenology is usually characterized by the formalism of closed quantum systems, where the evolution of the state, in vacuum, is fully described by a Hamiltonian
\begin{eqnarray}
 i\dfrac{d}{dt} \nu_{j} = \mathcal{H}\nu_{j}.
 \label{eq:schoe}
\end{eqnarray}
The $(\nu_{j})^T=(\nu_1,\nu_2,\nu_3)$ are the neutrino mass eigenstates and $\mathcal{H}$ is the Hamiltonian in mass basis, $\mathcal{H}={\rm diag}(\mathcal{H}_0, \mathcal{H}_0+\Delta m^2_{21}/2E, \mathcal{H}_0+\Delta m^2_{31}/2E)$, where $\mathcal{H}_0$ is a constant, not relevant for neutrino oscillation phenomenology, $\Delta m^2_{ij}\equiv m_i^2-m_j^2$ is the difference of the squared neutrino masses, with $i,j = 1,2,3$, and $E$ is the neutrino energy.
The solution of Eq.~(\ref{eq:schoe}) can be written as $\nu_j(t)=S_{ji}\nu_i(t=0)$, where $S$ is the evolution matrix of the neutrino system. Using the mixing matrix
$U$, which relates the flavor and the mass states, $\nu_{\alpha}=U_{\alpha j} \nu_j$, with $\alpha = e, \mu, \tau$, we can compute the neutrino probability as $P(\nu_{\alpha}\to \nu_{\beta}) \equiv  |(US^{\dagger}U^{\dagger})_{\beta\alpha}|^2$.

Due to the quantum nature of the neutrino evolution, when neutrinos are crossing large distances we may have decoherence effects induced by the separation of mass eigenstates~\cite{Akhmedov:2017mcc,Kersten:2015kio,Stankevich:2019zpf}. Here we will discuss a framework of decoherence and relaxation of neutrinos induced by their interaction with the environment, causing a change in the neutrino evolution. 
In the literature, there are different models for the interaction of a given system with the environment~\cite{Caldeira:1981rx,huang,Boriero:2017tkh,Cheng:2022lys}. For instance, at Reference~\cite{Caldeira:1981rx} the interaction is modeled as a set of harmonic oscillators. However, we will not restrict our analysis to a specific interaction model and will keep a phenomenological approach.

The general class of evolution of a given system due to environment interaction is called {\it open quantum system}. Assuming that neutrinos are described by such a system, we will discuss the implications of that in the neutrino oscillation framework testing it in present accelerator neutrino experiments. We will assume that the  neutrinos follow the Gorini-Kossakowski-Sudarshan-Lindblad equation in the mass basis~\cite{Gorini:1975nb,Lindblad1976}. Other  work formulates the decoherence and the relaxation scenarios in the flavor basis of neutrinos~\cite{Richter-Laskowska:2018ikv}. In the mass basis we have
\begin{eqnarray}
 \dfrac{d}{dt} \rho(t) = -i[\mathcal{H}, \rho(t)] + 
 \mathcal{D}[\rho(t)],
 \label{eq:lindblad.equation}
\end{eqnarray}
where $\rho$ and $\mathcal{H}$ are the density matrix and the Hamiltonian of the neutrino subsystem, respectively. $\mathcal{D}$ is an operator that has all the information to characterize the interaction of the neutrino subsystem with the environment, which can be described as 
\begin{eqnarray}
\mathcal{D}[\rho(t)] = \frac{1}{2}\sum_{\epsilon=1}^{N^2-1} \left( [V_{\epsilon}, \rho V^{\dag}_{\epsilon}] + [V_{\epsilon} \rho, V^{\dag}_{\epsilon}] \right)\label{eq:dissipation.term},
\end{eqnarray}
where  $V_{\epsilon}$ is a set of dissipative operators with the index $\epsilon$ going from 1 to $N^2-1$, and $N$ is the dimension of the SU($N$) group describing the interaction.

Considering the additional requirements of increasing Von Neumann entropy, probability conservation, complete positivity, and the decoherence and relaxation term $\mathcal{D}[\rho(t)]$, defined in neutrino mass basis, as described in Appendix~\ref{deco-prop}, we have the neutrino evolution matrix given by
\begin{eqnarray}
\dot{\rho}_{i} = \sum_j \mathcal{M}_{ij}\rho_j \quad \rm{and}
\quad  \rho_0=\sqrt{2/3},
\label{eq:lindblad.expanded.novo}
\end{eqnarray}
where the elements of the matrix $\mathcal{M}$ are
\begin{eqnarray}
\mathcal{M}_{ij}=\sum_k f_{ikj}\mathcal{H}_k+\mathcal{D}_{ij},
\label{eq:matrix.elements.M}
\end{eqnarray}
with $i,k,j =(1, \cdots, 8)$. The $\rho_{i}$ and $\mathcal{H}_i$ are, respectively, the $\rho$ and $\mathcal{H}$ projection in the SU(3) basis, $f_{ikj}$ are SU(3) structure constants and  $\mathcal{D}$ is the matrix defined by Eq.~(\ref{eq:dissipation.term}). The explicit format of the elements $\mathcal{D}_{ij}$ of the matrix $\mathcal{D}$ 
are computed on Appendix~\ref{deco-prop} and given by Eq.~(\ref{eq:matriz.dissipacao3}).

\subsection{Decoherence and relaxation scenarios}
\label{our-scenario}

The requirement of complete positivity stipulates that all eigenvalues of $\mathcal{D}$ must be negative, otherwise, the system would have abnormal behavior such as probabilities above one~\cite{Benatti:2000ph}.
For a diagonal matrix, 
\begin{eqnarray}
\mathcal{D} = \text{diag}\{\mathcal{D}_{11},\mathcal{D}_{22},\mathcal{D}_{33},\mathcal{D}_{44},\mathcal{D}_{55},\mathcal{D}_{66},\mathcal{D}_{77},\mathcal{D}_{88}\},
\label{eq:dissipation.matrix}
\end{eqnarray} 
the positivity condition is automatically satisfied if the diagonal elements are $\mathcal{D}_{ii} \le 0$. An additional condition is made in the literature in case there is energy exchange between the environment and the neutrino system, as discussed in Appendix~\ref{deco-prop}. 

The form of the matrix $\mathcal{M}$ that rules the 
neutrino evolution equation (Eqs.~(\ref{eq:lindblad.expanded.novo},~\ref{eq:matrix.elements.M})) is
\begin{eqnarray}
 \mathcal{M}  & = & \left(\begin{array}{cccccccc} 
  \mathcal{D}_{11} & -\Delta_{21} & 0 & 0 & 0 & 0 & 0 & 0 \\ 
  \Delta_{21} & \mathcal{D}_{22} & 0 & 0 & 0 & 0 & 0 & 0 \\
  0 & 0 & \mathcal{D}_{33} & 0 & 0 & 0 & 0 & 0 \\
  0 & 0 & 0 & \mathcal{D}_{44} & -\Delta_{31} & 0 & 0 & 0 \\
  0 & 0 & 0 & \Delta_{31} & \mathcal{D}_{55} & 0 & 0 & 0 \\
  0 & 0 & 0 & 0 & 0 & \mathcal{D}_{66} & -\Delta_{32} & 0 \\
  0 & 0 & 0 & 0 & 0 & \Delta_{32} & \mathcal{D}_{77} & 0 \\
  0 & 0 & 0 & 0 & 0 & 0 & 0 & \mathcal{D}_{88}  \end{array}
\right), 
\label{eq:matrix.M}
\end{eqnarray}
where $\Delta_{ij}=\Delta m_{ij}^2/2E$ and $\mathcal{D}_{ii}$ are the non-zero diagonal elements of the matrix $\mathcal{D}$. The solution of Eq.~(\ref{eq:lindblad.expanded.novo}), using the explicit formula for $\mathcal{M}$, is solved in the Appendix~\ref{deco-prop}. The full probability is  
\begin{eqnarray}
  P(\nu_{\alpha} \to \nu_{\beta})  & = & 
 \delta_{\alpha\beta}   
 -\sum_{j>i}\left\{4\mathbb{R}[{\rm W}_{\alpha\beta}^{ij}]
\left[\sin^2 \left( \frac{\Omega_{ij}}{4}L\right)\right]
-2  \left[\mathbb{I}[{\rm W}_{\alpha\beta}^{ij}]\right]
\sin\left( \frac{\Omega_{ij}}{2}L \right)
\right\}  e^{-\Gamma_{ij}L}
  \nonumber \\
 & & -2 \sum_{j>i}\left\{ \left[
 \frac{-\mathbb{R}[{\rm Y}_{\alpha\beta}^{ij}]
 \left(\Delta \mathcal{D}\right)_{ij}  + 
 \mathbb{I}[{\rm W}_{\alpha\beta}^{ij}]\left( 2\Delta_{ij}-\Omega_{ij}\right)}{\Omega_{ij}} \right]
\sin\left( \frac{\Omega_{ij}}{2}L \right)\right\}
 e^{-\Gamma_{ij}L} \nonumber \\
\!\!\!\!\!\!\!\!\!\!\!\!\!\! & &      -\frac{1}{2}\left\{\left(\frac{ 1 - 3|U_{\alpha 3}|^2}{\sqrt{3}} \right)\left(\frac{ 1 - 3|U_{\beta 3}|^2}{\sqrt{3}} \right)\left(1-e^{\mathcal{D}_{88}L}\right)  \right\} \nonumber \\
     & & -\frac{1}{2} \left\{ \left( |U_{\alpha 1}|^2 - |U_{\alpha 2}|^2 \right)\left( |U_{\beta 1}|^2 - |U_{\beta 2}|^2 \right) \left(1- e^{\mathcal{D}_{33}L}\right)\right\},
 \label{eq:prob.transicao.3.sabores.osc.dec_termos_U}
\end{eqnarray}
where ${\rm W}_{\alpha\beta}^{ij}\equiv U^{*}_{\alpha i}U_{\alpha j}U_{\beta i}U^{*}_{\beta j}$ is the Jarlskog invariant~\cite{Jarlskog:1985ht,Jarlskog:1985cw} and ${\rm Y}_{\alpha\beta}^{ij}\equiv U^{*}_{\alpha i}U_{\alpha j}U^{*}_{\beta i}U_{\beta j}$ is a new amplitude that appears in the decoherence scenario. This later amplitude is not invariant by Majorana phases, as noticed before in Ref.~\cite{Benatti:2001fa,Oliveira:2010zzd,Capolupo:2018hrp,Buoninfante:2020iyr}. The quantities $\Gamma_{ij}$ and $\Omega_{ij}$ 
are given in Eq.~(\ref{eq:gamma.sector3}) and (\ref{eq:gamma.sector1}) of Appendix~\ref{deco-prop}. In the limit of null decoherence and relaxation we have, $\Omega_{ij} \to 2\Delta_{ij}, \mathcal{D}_{ij}\to 0, \Gamma_{ij}\to 0,\left(\Delta \mathcal{D}\right)_{ij}\to 0$, with all the terms in the first line of Eq.~(\ref{eq:prob.transicao.3.sabores.osc.dec_termos_U}) recovering the usual three neutrino oscillation, while the terms in the other lines vanish.

The oscillation probability shown in Eq.~(\ref{eq:prob.transicao.3.sabores.osc.dec_termos_U}) has damping terms, which appear in:
\begin{enumerate}
\item the oscillatory term, shown in the first and second lines of Eq.~(\ref{eq:prob.transicao.3.sabores.osc.dec_termos_U}), which is governed by the $\Gamma_{ij}$ parameters. This is usually called \emph{decoherence} in the literature~\cite{GUZZO2016408};
\item the non-oscillatory term, in the third and fourth lines of Eq.~(\ref{eq:prob.transicao.3.sabores.osc.dec_termos_U}). This phenomenon is referred to as relaxation in the literature~\cite{GUZZO2016408}.
\end{enumerate}

From our choice of decoherence and relaxation matrix $\mathcal{D}$ and the $2\times 2$ block-diagonal nature of $\mathcal{M}$, we observe that different sub-matrices will decouple in the evolution and in the neutrino probability as well. For instance, the elements $\mathcal{D}_{11}$ and $\mathcal{D}_{22}$ are correlated to the \emph{solar neutrino oscillation} (which is guided by $\Delta m_{21}^2$), while $\mathcal{D}_{44}$, $\mathcal{D}_{55}$, $\mathcal{D}_{66}$, and $\mathcal{D}_{77}$ have correlation to the \emph{atmospheric/long-baseline neutrino oscillation} (which is related to $\Delta m_{31}^2$ and $\Delta m_{32}^2$). In other words, the oscillation that is mostly between the first and second generation, i.e. $i,j=1,2$, implies that the main role of the decoherence will be made by the $\mathcal{D}_{11}$ and $\mathcal{D}_{22}$ and then the more important terms are $\Gamma_{21}$ and $\Omega_{21}$.

Next, we will describe the different decoherence and relaxation scenarios that we are going to investigate. Considering that we have eight diagonal elements, $\mathcal{D}_{ii}$, and using their explicit form given by Eq.~(\ref{eq:matriz.dissipacao3}), we should find a self-consistent solution for $\mathcal{D}$ in terms of the requirements of strict increase of entropy, probability conservation and complete positivity. We then decide to investigate three possible scenarios, described below, and summarized in Table~\ref{tab:decoh.cases}:
\begin{enumerate}
\item Case 1: We choose a democratic scenario, where all entries $\mathcal{D}_{ii}$ are non-zero and equal, $\mathcal{D}_{ii} = -\Gamma$, for $i = 1, \cdots, 8$. Under these assumption, we obtain $\Gamma_{ij}~\to~\Gamma$, $\Omega_{ij}\to 2\Delta_{ij}$, and $(\Delta \mathcal{D})_{ij}~\to~0$ and the second line of the oscillatory term in Eq.~(\ref{eq:prob.transicao.3.sabores.osc.dec_termos_U}) is vanished. In this case, we have decoherence and relaxation at the same time.

\item  Case 2: We consider {\em no energy exchange} (see Appendix~\ref{deco-prop} for details), implying that $\mathcal{D}_{33}=\mathcal{D}_{88}=0$, with all others elements $\mathcal{D}_{ii}= -\Gamma$. Obviously, this will also result in $\Gamma_{ij} \to \Gamma$, $\Omega_{ij}\to 2 \Delta_{ij}$, and $(\Delta \mathcal{D})_{ij}\to 0$, vanishing the second, third and fourth lines of Eq.~(\ref{eq:prob.transicao.3.sabores.osc.dec_termos_U}). The only difference of the resulting oscillatory term (first line) of the probability to the standard oscillation probability is the exponential damping terms.
In this case, we have decoherence only.

\item  Case 3: The difference between this case and Case 2 is that we will assume that the  effect of the decoherence will be happening in the 
$\mathcal{D}$ sector relevant for long-baseline experiments only. This implies that $\mathcal{D}_{11} = \mathcal{D}_{22} = 0$. Thus we continue assuming {\em no energy exchange},  $\mathcal{D}_{33}=\mathcal{D}_{88}=0$, and the other elements are $\mathcal{D}_{ii}\to -\Gamma$, for $i=4,\cdots ,7$. The probability is the same as in Case 2 except for the absence of the exponential term for $\Gamma_{21}$. In other words, we keep the terms only in the \emph{atmospheric/long-baseline neutrino oscillation}, related to $\Delta m_{31}^2$ and $\Delta m_{32}^2$ mass scales. In this case, we have decoherence only.
\end{enumerate}

\begin{table}[!htp]
\centering
\begin{tabular}{c|cccccccccc}
{\rm Models \hspace{0.1 mm}} & $\mathcal{D}_{11}$ & $\mathcal{D}_{22}$ & $\mathcal{D}_{33}$ & $\mathcal{D}_{44}$ & $\mathcal{D}_{55}$ & $\mathcal{D}_{66}$ & $\mathcal{D}_{77}$ & $\mathcal{D}_{88}$ & $\Gamma_{21}$ & $\Gamma_{31}=\Gamma_{32}$ \\
\hline
{\rm Case 1 \hspace{-0.1 mm} } & -$\Gamma$ & -$\Gamma$ & -$\Gamma$ & -$\Gamma$ & -$\Gamma$ & -$\Gamma$ & -$\Gamma$ & -$\Gamma$ & $\Gamma$ & $\Gamma$   \\
{\rm Case 2 \hspace{-0.1 mm} } & -$\Gamma$ & -$\Gamma$ & 0 & -$\Gamma$ & -$\Gamma$ & -$\Gamma$ & -$\Gamma$ & 0 & $\Gamma$ & $\Gamma$   \\
{\rm Case 3 \hspace{-0.1 mm} } & 0 & 0 & 0 & -$\Gamma$ & -$\Gamma$ & -$\Gamma$ & -$\Gamma$ & 0 & 0  & $\Gamma$   \\
\end{tabular}
\caption{The decoherence and relaxation parameters that characterize the models that we investigate.}
\label{tab:decoh.cases}
\end{table}

From Table~\ref{tab:decoh.cases} we easily note that in all cases $\Gamma_{31} = \Gamma_{32}$, and they are equal to $\Gamma$. In Case 1, we have $\Gamma_{21} = \Gamma$ and relaxation is allowed (but constrained to the same value of $\Gamma$). In Case 2, we also have $\Gamma_{21} = \Gamma$, but no relaxation is allowed. And in Case 3, we set $\Gamma_{21} = 0$ and no relaxation is allowed also. Thus, these scenarios allow us to compare cases 1 and 2 to investigate any effect due to relaxation. And the comparison of cases 2 and 3 allows the investigation of not constraining the solar and atmospheric sectors to the same decoherence parameter.

\subsection{Energy dependence}
\label{energy-dependence}

The energy dependence of the decoherence and relaxation parameter does not have a precise underlying theory. In the literature there are different proposals of which we can cite as examples the following: $(i)$ energy independent, $E^0$, $(ii)$ $E^2$ dependence, appealing to {\it quantum gravity} arguments~\cite{Barenboim:2004wu}, and $(iii)$ $E^{-1}$ dependence, assuming to have similar dependence of usual oscillation phase. Thus, in general, we can consider an energy dependence like $E^n$ and write the decoherence and relaxation parameter as \cite{Farzan:2008zv,Lisi:2000zt},
\begin{eqnarray}
\Gamma = \gamma_0 \left(\frac{E}{E_0}\right)^n,
\label{eq:Gamma.phenomenology}
\end{eqnarray}
where $\gamma_0$ is the constant parameter, $E$ is the neutrino energy, $n$ is the power-law dependence, and $E_0$ is an energy reference that we set as a constant and equal to $1$~GeV. In the following section we are going to analyze the three different cases listed in Table~\ref{tab:decoh.cases}, for three different power-law dependence, $n = -2, 0, +2$, to constrain these cases using all available information from the MINOS~\cite{Adamson:2013whj} and the T2K~\cite{Abe:2017bay,Abe:2017uxa} experiments.  The choice to investigate such values of $n$ encloses the scenarios $n=\pm 1$ as intermediate values between $n=0$ and $n=\pm 2$.

\begin{figure}[hbt]
\hspace{-0.6cm}
\includegraphics[scale=0.31]{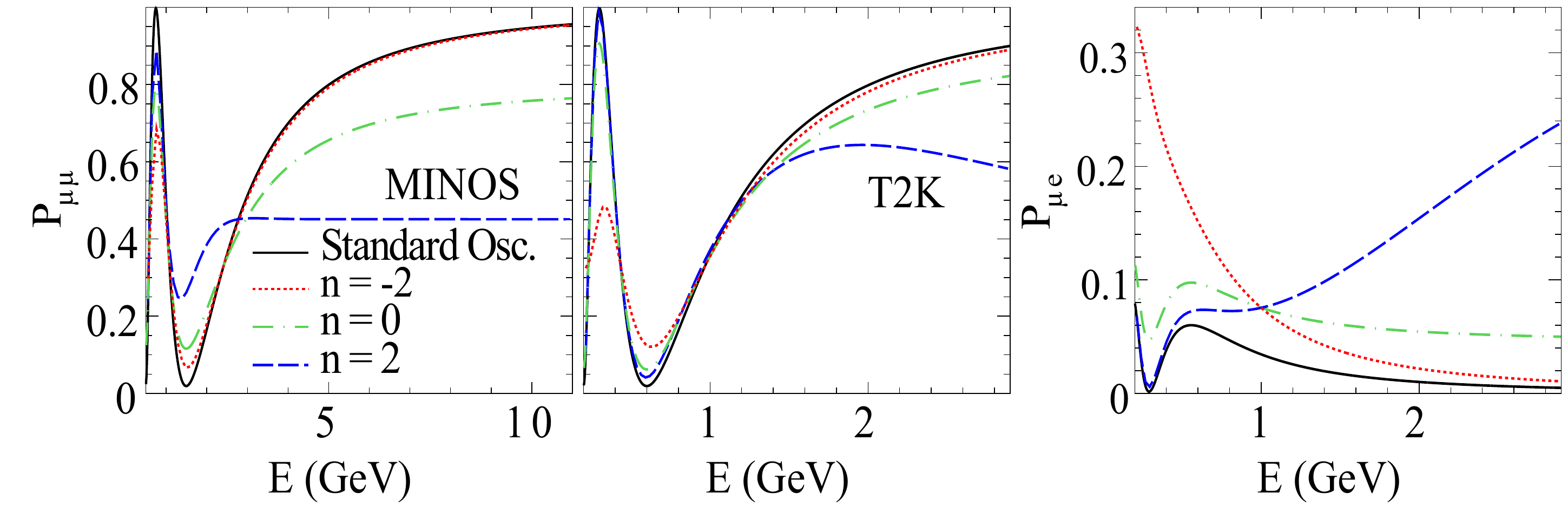}
\caption{The survival probability, $P_{\mu\mu}\equiv P(\nu_{\mu}\to \nu_{\mu})$, for MINOS (left) and T2K (middle), and the transition probability, $P_{\mu e}\equiv P(\nu_{\mu}\to \nu_{e})$, for T2K (right), as a function of energy. We show the probability curves for the standard oscillation model (black solid) and for the decoherence and relaxation model, with $n = -2$ (red dotted), $n = 0$ (green dashed-dotted), and $n = +2$ (blue dashed). The parameter $\gamma_0$ is fixed and equal to $10^{-22}$~GeV.}
\label{effect.of.gamma_0}
\end{figure}

The left and middle panel of Figure~\ref{effect.of.gamma_0} show the survival probability for MINOS and T2K, respectively, while the right one shows the transition ($\nu_\mu \to \nu_e$) probability for T2K.
To present the behavior of the survival and transition probabilities under the decoherence and relaxation framework we choose, as an example, the Case 1  (Table~\ref{tab:decoh.cases}), for different values of $n$. We used the following oscillation parameters to be fixed to the best-fit values of Ref.~\cite{Esteban:2018azc}, $\sin^2\theta_{23} = 0.580$, $\sin^2\theta_{12} = 0.310$, $\sin^2\theta_{13} = 0.02240$, $\Delta m^2_{31} =2.525\times10^{-3}~\text{eV}^2$, $\Delta m^2_{21}~=~ 7.39\times10^{-5}~\text{eV}^2$, and $\delta_{\rm CP}=217^{\circ}$. And to investigate the effect of the decoherence and relaxation we set $\gamma_0=10^{-22}$~GeV. 

Comparing the decoherence and relaxation probabilities to the standard oscillation probability, shown in Figure~\ref{effect.of.gamma_0}, we observe that, for a certain value of $\gamma_0$, the effect on muon neutrino survival probability in MINOS for $n=+2$ is stronger than for $n=-2$. On the other hand, the $n=+2$ for muon neutrino survival probability in T2K is very close to the standard oscillation curve for energies below 1.5~GeV (relevant for T2K disappearance analysis), and no significant effect is noted.
For the muon to electron neutrino conversion probability, shown in the right panel of Figure~\ref{effect.of.gamma_0}, we see that the probabilities including decoherence and relaxation are always higher than the standard oscillation case. But for energies below 1~GeV, which is the relevant energy range for the T2K $\nu_e$ appearance analysis, the effect for $n=-2$ is stronger than the effect of other values of $n$.

We then observe two clear domains: below and above 1 GeV, where depending on the energy range of the experiment we can better constrain positive or negative values of $n$. Since the energy range of MINOS is totally above 1~GeV, we expect a stronger constraint on $\gamma_0$ for $n = +2$ than for $n = -2$. For T2K, the energy spectrum is both below and above 1~GeV, therefore we expect similar constraints on $\gamma_0$ for the considered values of $n$. But since MINOS energies are higher than T2K energies, the $n =+2$ constraint from MINOS is expected to be more stringent than the one from T2K. This complementary behavior between MINOS and T2K makes their combination interesting to impose constraints on $\gamma_0$ for both negative and positive values of $n$. Summarizing, MINOS (T2K) would imply a more stringent constraint on $\gamma_0$ for $n=+2$ ($n=-2$) than for the other considered values.

\section{Dataset and Fitting Procedure}
\label{sec:analysis}

We have performed an analysis using MINOS~\cite{Adamson:2013whj} and T2K~\cite{Abe:2017bay,Abe:2017uxa} published data. MINOS experiment used two detectors, located at 1~km and 735~km from the target, exposed to a neutrino beam produced at FERMILAB. Its beam-line could be configured to optimize muon neutrino or anti-neutrino composition. In this analysis we used both neutrino and anti-neutrino disappearance data~\cite{Adamson:2013whj} from the neutrino optimized configuration, which comprised 10.71 $\times$ $10^{20}$ POT (protons on target). T2K is a 295~km baseline experiment consisted of two detectors exposed to a neutrino beam produced at J-PARC. The T2K neutrino beam has also two configurations: neutrino and anti-neutrino runs. However, differently from MINOS, T2K does not distinguish neutrino and anti-neutrino events. The T2K dataset we used are from $\nu_\mu$ disappearance and $\nu_e$ appearance analyses from both neutrino ($7.48 \times 10^{20}$ POT) and anti-neutrino ($7.47 \times 10^{20}$ POT) runs~\cite{Abe:2017bay, Abe:2017uxa}.

Due to the number of events per energy bin $i$ in MINOS data, we used the following Gaussian $\chi^2$
\begin{eqnarray}
\chi^2_{\rm MINOS} &=& \sum_i \left( \frac{N^{\rm th}_i - N^{\rm d}_i}{\sigma_i } \right)^2,
\label{eq:chi2.gaussian}
\end{eqnarray}
where the number of data events is $N^{\rm d}_i$, the total error is $\sigma_i$, and the prediction of the theoretical model is $N^{\rm th}_i = (1 + \alpha) N^{\rm sig}_i + (1 + \beta) N^b_i$, which considered the signal, $N^{\rm sig}_i$, and background, $N^b_i$, contributions with normalization parameters, $\alpha$ and $\beta$, respectively.
Gaussian penalty terms were included in the $\chi^2$ for the normalization parameters with uncertainties $\sigma_\alpha = 14.7\%$ and $\sigma_\beta = 4.0\%$~\cite{Adamson:2007gu}.

For the T2K data analyses the calculations were performed with a $\chi^2$ given by
\begin{eqnarray}
\chi^2_{\rm T2K} &=& 2\sum_{\rm i} \left[ N^{\rm th}_{\rm i} - N^{\rm d}_{\rm i} - N^{\rm d}_i~{\rm ln} \left( \frac{N^{\rm th}_i}{N^{\rm d}_{\rm i}} \right) \right],
\label{eq:chi2.poisson}
\end{eqnarray}
where the theoretical prediction of events is
\begin{eqnarray}
N^{\rm th}_i = \left[ 1 + \alpha + t\left( \frac{E_i - \overline{E} }{E_{\rm max}} \right) \right] (N^{\rm sig}_i + N^b_i).
\end{eqnarray}
In addition to the normalization parameter $\alpha$ we introduced a term allowing a distortion of the energy spectrum~\cite{Fogli:2002pt,Huber:2002mx}, where the parameter $t$ is the tilt, $E_i$ is the average bin energy, $\overline{E}$ is the average spectrum energy, and $E_{\rm max}$ is the maximum energy of the spectrum. The uncertainties of the penalty terms for the normalization and tilt parameters were both set equal to 20\% (15\%) for the disappearance (appearance) analysis. The details of the analyses are discussed at the Appendix~\ref{app:validation}.

We have first validated our procedure by the $\chi^2$ analysis of each dataset as a function of $\sin^2 \theta_{23}, \sin^2 \theta_{13}, \delta_{\rm CP}$ and $\Delta m_{32}^2$, for the standard oscillation model, under the normal mass ordering (the $\Delta m_{31}^2$ parameter is given by $\Delta m_{31}^2\equiv\Delta m_{32}^2+ \Delta m_{21}^2$). The oscillation parameters  $\sin^2\theta_{12}~=~0.307$ and $\Delta m^2_{21}=7.54\times 10^{-5}~\text{eV}^2$ are fixed to the best-fit values from Ref.~\cite{Esteban:2016qun}. Our results agree reasonably well with the results of the official MINOS and T2K analyses. For the decoherence and relaxation model discussed in this study, there are two additional parameters, $\gamma_0$ and $n$, both defined in Eq.~(\ref{eq:Gamma.phenomenology}).

\begin{figure}[!htp]
\subfloat{\includegraphics[scale=0.4]{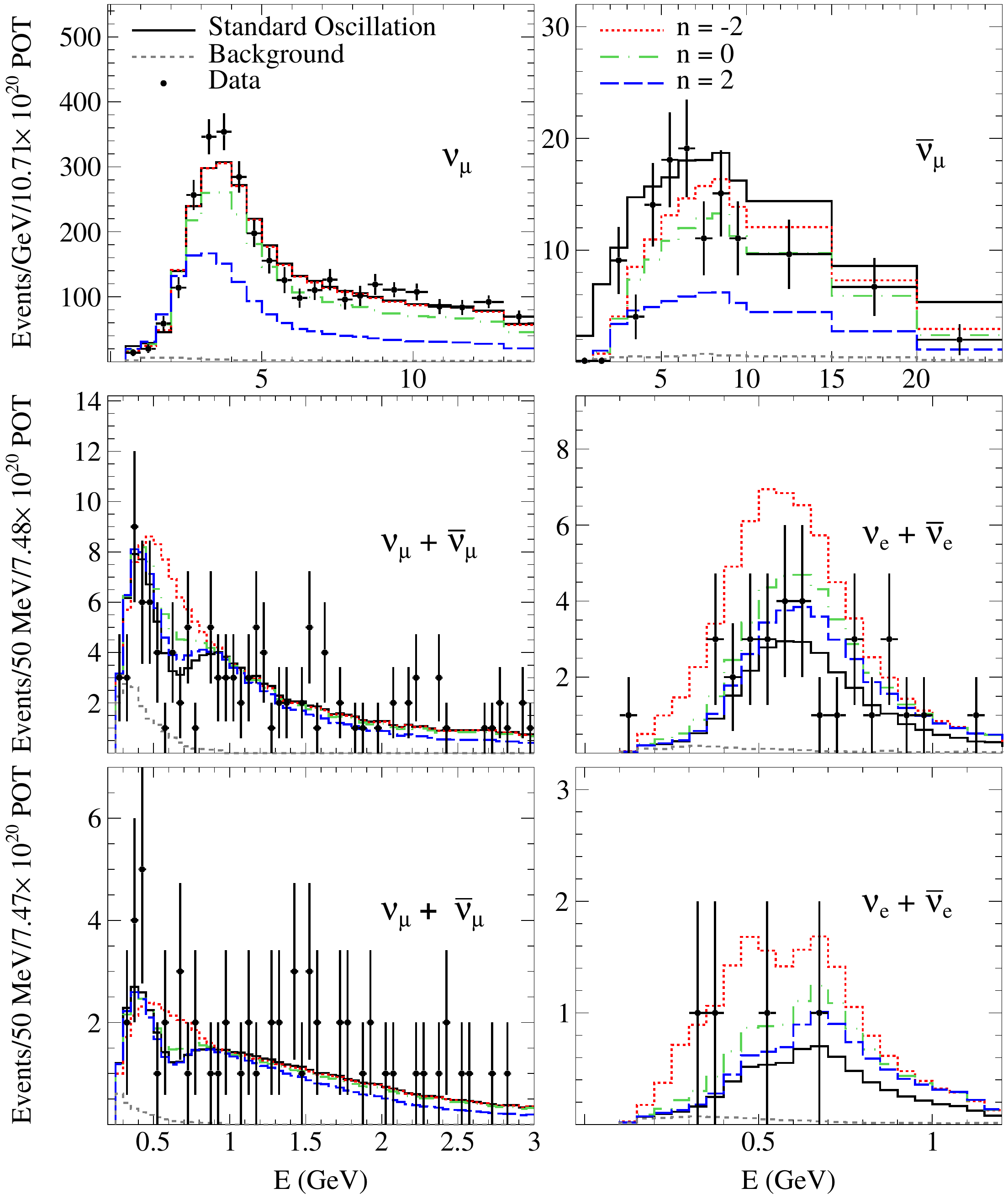}}
\caption{Top panel: spectra of MINOS data for $\nu_\mu$ (left) and $\overline{\nu}_\mu$ (right) disappearance analyses. Middle and Bottom panels: spectra of T2K data for disappearance (left) and appearance (right) analyses for the {\it neutrino mode} (middle) and {\it anti-neutrino mode} (bottom). The best-fit curve for the standard oscillation model (solid black curves), obtained on our validation process for each experimental dataset individually. The other curves were obtained for a fixed $\gamma_0 = 10^{-22}$~GeV value and for the three powers, $n = -2,0,+2$, given by the red dotted, green dashed-dotted, and blue dashed curves, respectively.}
\label{spectrum.effect.of.gamma_0.minos.t2k}
\end{figure}

The top panel of Figure~\ref{spectrum.effect.of.gamma_0.minos.t2k} shows the extracted spectra of neutrino events for MINOS disappearance analyses, while the middle (bottom) panel shows the T2K disappearance and appearance analyses for the neutrino (anti-neutrino) mode.
The solid curves presented in all spectra are the standard oscillation best-fit curves, obtained individually for each experiment and spectrum during our validation process. With the only purpose to observe the decoherence and relaxation effects on MINOS and T2K spectra we kept the best-fit parameters obtained for each experiment and included a $\gamma_0$ value equal to $10^{-22}$~GeV for different values of $n$.
This figure shows that MINOS is not sensitive for $n = -2$, while $n = +2$ has the more prominent effect. On the order hand, the T2K spectra show that $n = -2$ has a stronger effect than the other values of $n$ for all four data-sets. These observations are all in agreement with the previous discussion in Section~\ref{our-scenario}.

In addition to the analyses on MINOS and T2K data-set separately, we also performed a combined analysis with $\chi^2=\chi^2_{\rm MINOS} +\chi^2_{\rm T2K}$. On that analysis we investigate the effect of including a reactor constraint on $\sin^2 2\theta_{13}$, where we used a Gaussian $\chi^2$ shape based on the result from Ref.~\cite{An:2016ses},
\begin{eqnarray}
\chi^2_{\text{\rm reactor}} = \left(\frac{\sin^2 2\theta_{13} - 0.0841}{0.0033}\right)^2\label{eq:chi-square.reactor.constrain}.
\end{eqnarray}

\section{Results}
\label{sec:results}

\begin{figure}[htb]
\centering
\includegraphics[width=1.\textwidth]{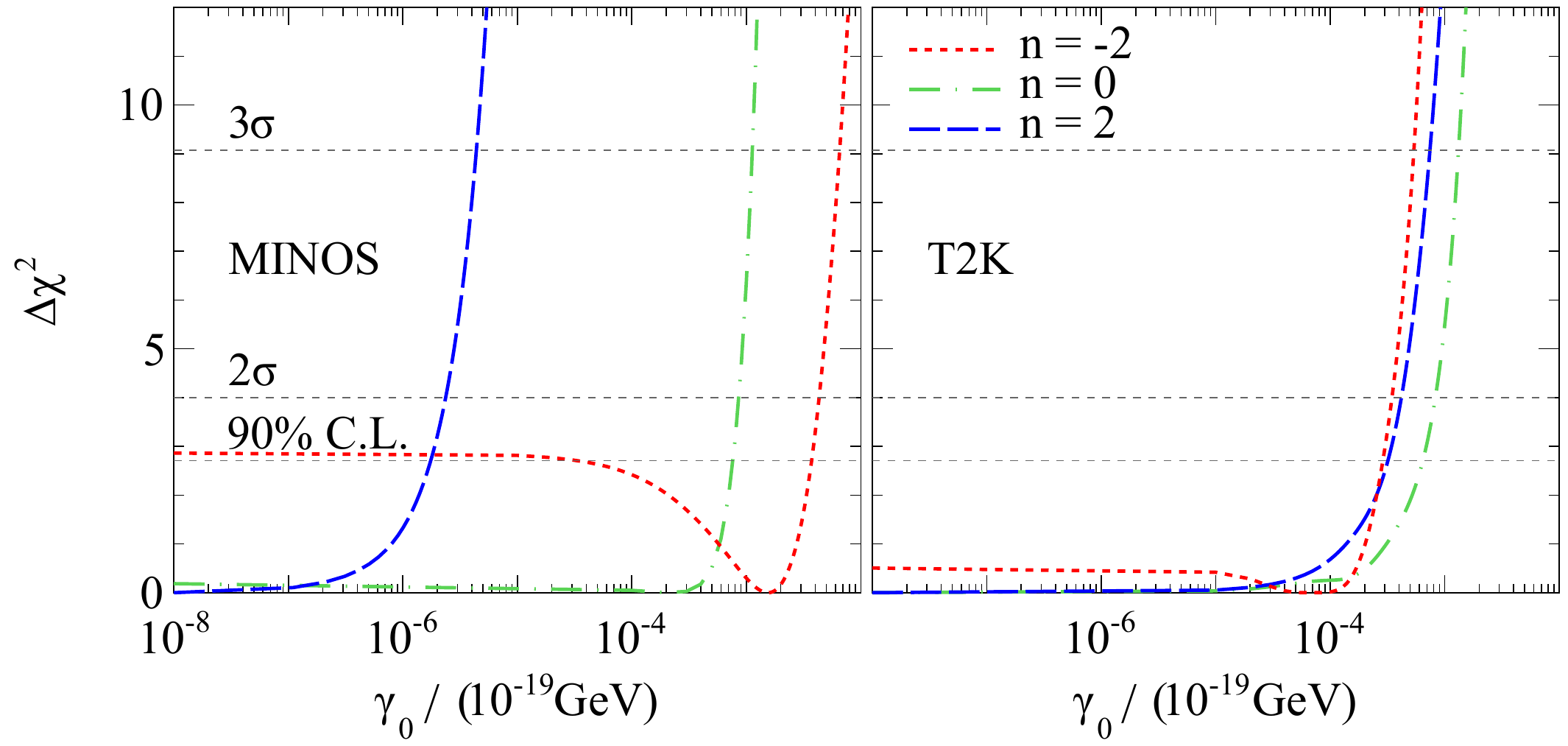} \hfill
\vspace{-0.8cm}
\caption{Projection of  the $\Delta\chi^2\equiv \chi^2 - \chi^2_{\rm min}$ as a function of $\gamma_0$ parameter. The left~(right) panel shows the bounds obtained for the analysis of MINOS~(T2K) data for $n=-2,0,2$.}
\label{fig:1d.gamma_0.c1.minos.t2k}
\end{figure}

The results presented here comprise the analyses of all the decoherence and relaxation models introduced at Table~\ref{tab:decoh.cases}, using the MINOS and T2K (separately and combined) dataset. We also present an investigation of the effect of a reactor constraint and compare our upper bounds on $\gamma_0$ with previous bounds. For all analyses performed we consider as free variables the oscillation parameters described before, $\sin^2 \theta_{23}$, $\sin^2 \theta_{13}$, $\Delta m_{32}^2$ and $\delta_{\rm CP}$, and the decoherence and relaxation parameters, $\gamma_0$ and $n$. The solar sector neutrino oscillation parameters, $\sin^2 \theta_{12}$ and $\Delta m_{21}^2$, are kept fixed and we consider the normal mass hierarchy only. We scan all these free parameters to find the best-fit solution and the allowed regions for a given scenario, i.e., a combination of one of the cases and one of the values of $n$.

\subsection{Individual and combined analyses on MINOS and T2K dataset}

We first show the projection of $\Delta \chi^2\equiv \chi^2 - \chi^2_{\rm min}$ as a function of the $\gamma_0$ parameter, for MINOS and T2K analyses, separately, in Figure~\ref{fig:1d.gamma_0.c1.minos.t2k}. It is shown the curves for Case 1 only, since we obtained similar behavior for the curves of the others investigated cases. The horizontal lines present the $\chi^2$ values for certain confidence levels, considering one degree of freedom. The left (right) panel presents the bounds obtained by MINOS (T2K) for $n=-2,0,+2$, given by the red dotted, green dashed-dotted and blue dashed curves, respectively.

In agreement to the expectation discussed before, the result for MINOS shows a better constraint on $\gamma_0$ for $n=+2$ than for the other values of $n$. For $n=-2$ we have found a bound two orders of magnitude less stringent than for $n = +2$, and a global minimum different from zero, with a significance of about 90$\%$ C.L. On the other hand, the analysis for T2K data shows similar constraints on $\gamma_0$ for $n=-2$ and $n=+2$, with the weaker bound obtained for $n=0$. Based on the discussion of Figure~\ref{spectrum.effect.of.gamma_0.minos.t2k}, this result is explained by the fact that the T2K data is dominated by the $\nu_\mu + \overline{\nu}_\mu$ disappearance spectra, which has sensitivity for both $n=-2$ and $n=+2$. Despite the spectra of $\nu_{e} + \overline{\nu}_{e}$ appearance presenting a major effect for $n=-2$ energy dependence (for neutrino energies below 1~GeV), the poor statistics from these samples does not significantly improve the limits with regard to the analyses for $n = 0$ and $+2$. It is worth mentioning from Figure~\ref{fig:1d.gamma_0.c1.minos.t2k} that some of the scenarios for Case 1, on both MINOS and T2K dataset, result in a best-fit value of $\gamma_0$ different from zero. Such behavior, which is also present on the other cases, can potentially effect the best-fit values and allowed regions of the neutrino oscillation parameters, as we will see later.

\begin{table}[htb]
\footnotesize
\centering
\tabcolsep=0.11cm
\begin{tabular}{|c|@{\hspace{2mm}}c@{\hspace{3mm}}c@{\hspace{3mm}}c@{\hspace{2mm}}|}
\hline
& $n=-2$ &  $n=0$ & $n=2$\\
\hline
\textbf{MINOS (this work)} & & &\\
\hline
Case 1 ($\Gamma_{31} = \Gamma_{32} = \Gamma_{21}$, with relaxation)
& $(0.33 - 37.0)\times10^{-23}$ & $6.8\times10^{-23}$ & $1.7\times10^{-25}$\\
Case 2 ($\Gamma_{31} = \Gamma_{32} = \Gamma_{21}$, no relaxation)
& $30.0\times10^{-23}$          & $6.5\times10^{-23}$ & $2.4\times10^{-25}$\\
Case 3 ($\Gamma_{31} = \Gamma_{32}$, $\Gamma_{21} = 0$, no relaxation)
& $19.0\times10^{-23}$          & $5.9\times10^{-23}$ & $2.5\times10^{-25}$\\
\hline
\textbf{T2K (this work)} & & &\\
\hline
Case 1 & $2.8\times10^{-23}$ & $6.2\times10^{-23}$ & $3.1\times10^{-23}$\\
Case 2 & $2.9\times10^{-23}$ & $5.2\times10^{-23}$ & $3.3\times10^{-23}$\\
Case 3 & $1.7\times10^{-23}$ & $3.9\times10^{-23}$ & $4.1\times10^{-23}$\\
\hline
\textbf{MINOS+T2K (this work)} & & &\\
\hline
Case 1 & $2.9\times10^{-23}$ & $6.6\times10^{-23}$ & $2.3\times10^{-25}$\\
Case 2 & $3.4\times10^{-23}$ & $6.1\times10^{-23}$ & $2.9\times10^{-25}$\\
Case 3 & $2.0\times10^{-23}$ & $5.0\times10^{-23}$ & $3.3\times10^{-25}$\\
\hline

\textbf{MINOS+T2K+RC (this work)} & & &\\
\hline
Case 1 & $2.7\times10^{-23}$ & $6.4\times10^{-23}$ & $2.3\times10^{-25}$\\
Case 2 & $3.2\times10^{-23}$ & $6.5\times10^{-23}$ & $2.8\times10^{-25}$\\
Case 3 & $1.7\times10^{-23}$ & $4.8\times10^{-23}$ & $3.3\times10^{-25}$\\
\hline
\textbf{Previous Bounds} & &  & \\
\hline
Ref.~\cite{Lisi:2000zt} & -- & $3.5\times10^{-23}$ & $9.0\times10^{-28}$\\
Ref.~\cite{deOliveira:2013dia} & $ 2.0\times 10^{-22}$ & $(0.6 - 5.5)\times 10^{-23}$ & $5.0\times 10^{-25}$\\
Ref.~\cite{Gomes:2016ixi} & -- & $6.8\times 10^{-22}$ & -- \\

Ref.~\cite{Coloma:2018idr} (a) & $2.8\times10^{-18}$ & $4.0\times10^{-24}$ & $1.0\times10^{-31}$\\
Ref.~\cite{Coloma:2018idr} (b) & $4.3\times10^{-20}$ & $8.2\times10^{-23}$ & $1.1\times10^{-25}$\\
\hline
\textbf{Sensitivity} & &  &  \\
\hline 
Ref.~\cite{Gomes:2018inp} (c) & -- & $4.7\times 10^{-24}$ &  -- 
\\
Ref.~\cite{Gomes:2018inp} (d) & -- & $7.7\times 10^{-25}$ &  -- \\
\hline
\end{tabular}
\caption{Our bounds on $\gamma_0$, at 90\% C.L. (1 degree of freedom), from the data analyses for MINOS only, T2K only, combined MINOS+T2K, and combined MINOS+T2K with reactor constraint. Previous bounds based on phenomenological analyses of published data (Ref.~\cite{Lisi:2000zt} at 90$\%$ C.L. for Super-Kamiokande, Ref.~\cite{deOliveira:2013dia} at 68$\%$ C.L. for MINOS, ~Ref.~\cite{Gomes:2016ixi} at 95\% C.L. for KamLAND, and Ref.~\cite{Coloma:2018idr} (a) and (b), at 95$\%$ C.L., for IceCube and DeepCore, respectively) and on sensitivity analyses (Ref.~\cite{Gomes:2018inp} (c) and (d), at 90\% C.L., for DUNE under two different flux configurations). All bounds are in GeV.}
\label{table:summary-tableNO}
\end{table}

At Table~\ref{table:summary-tableNO} we present the bounds on $\gamma_0$ parameter, at the 90\% C.L., obtained by the individual analyses of MINOS and T2K, for all the cases and the different values of $n$ considered in this study. We observe that for each $n$ and dataset (MINOS or T2K) there is no significant difference between the cases (1, 2, and 3). Indeed, none of those differences is greater by a factor of 2 than the others. This independence of the case is a hint that neither of the experimental dataset used has sensitivity for the relaxation effect (comparing the cases 1 and 2) or the constraint effect between the solar and the atmospheric sectors (comparing the cases 2 and 3).

The individual analyses reported at Table~\ref{table:summary-tableNO} also show that for $n = +2$ the MINOS results are two orders of magnitude more stringent than the T2K results. While for $n = -2$ the T2K results are one order of magnitude more stringent than the MINOS results, in a clear manifestation of the complementary behaviour between the two datasets. For $n = 0$ all the results are very similar between MINOS and T2K. And as already mentioned, all these observations are independent of the case investigated.

\begin{figure}[ht]
\centering
\includegraphics[width=1.\textwidth]{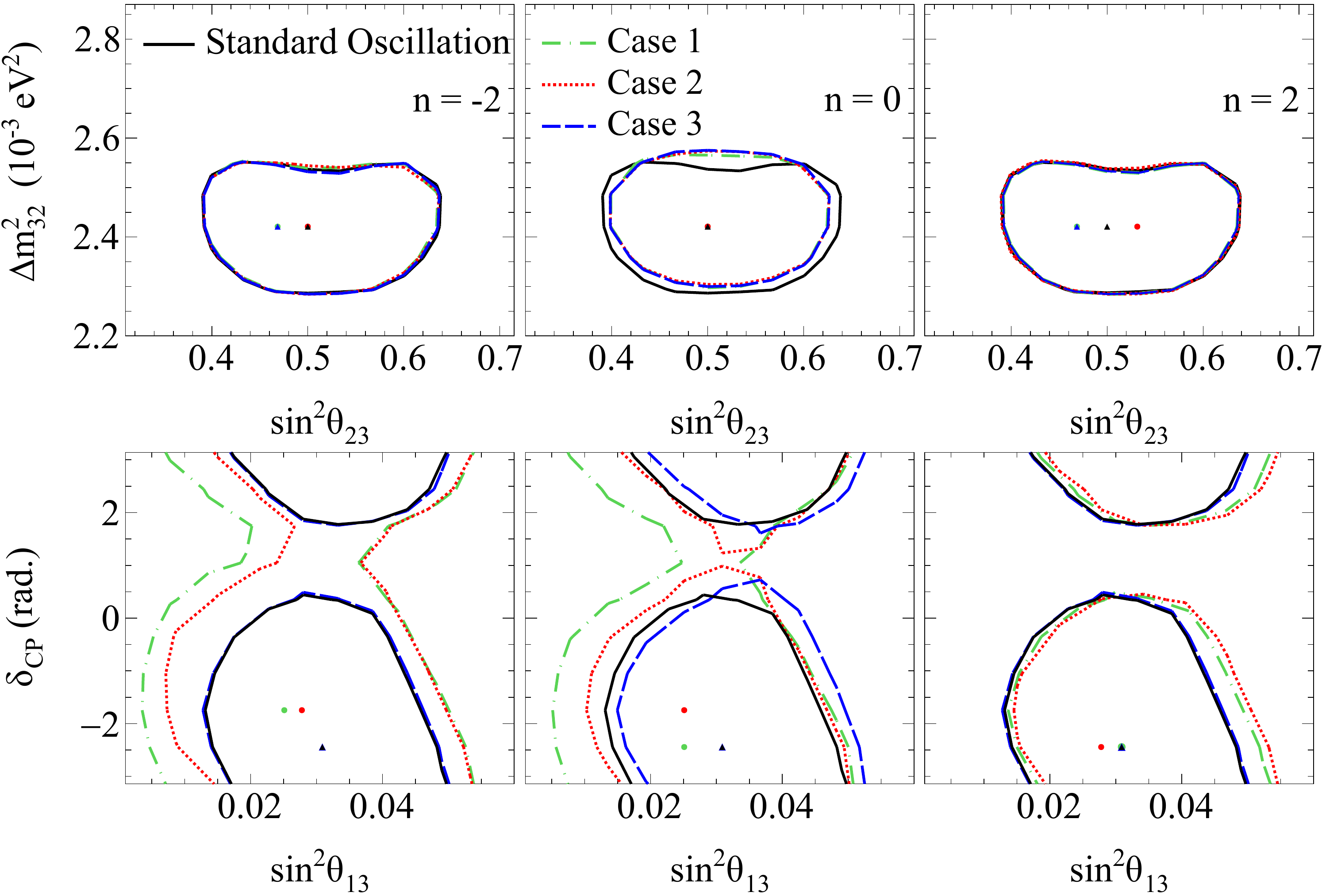}
\caption{The allowed regions of oscillation parameters are presented at 90$\%$ C.L. for the Case 1 (green dashed-dotted), Case 2 (red dotted), Case 3 (blue dashed) and the Standard Oscillation (black solid) for MINOS+T2K analysis. Following the columns from the left to right we have $n=-2,0,+2$, respectively. Top panel: The projections of $\Delta m^2_{32} - \sin^2\theta_{23}$. Bottom panel: The projections of $\delta_{\rm CP}-\sin^2\theta_{13}$.  The best-fit values for each analysis are shown, by the red circle, green circle and blue triangles, respectively. }
\label{fig:2d.dm32.s23.s13.dcp.c1.c2.c3.minos.plus.t2k}
\end{figure}

A combined analysis of these two complementary dataset, with regard to the models we investigate, could give us the best of each experiment to place bounds on the decoherence and relaxation scenarios. The Figure~\ref{fig:2d.dm32.s23.s13.dcp.c1.c2.c3.minos.plus.t2k} shows the best-fit values and the allowed regions, at 90$\%$ C.L., of the oscillation parameters for the cases 1, 2 and 3. The standard oscillation scenario, given by the black solid curve, is also presented. The left, middle, and right columns show the results for $n = -2$, $0$, and $+2$, respectively.

 There is no significant effect of the decoherence and relaxation models on the standard oscillation parameters, as we can see from Figure~\ref{fig:2d.dm32.s23.s13.dcp.c1.c2.c3.minos.plus.t2k}. The top panel of this figure presents the $\Delta m^2_{32} - \sin^2\theta_{23}$ projections, from where we do observe that for some scenarios the inclusion of the decoherence moves the best-fit value to $\sin^2 \theta_{23}\neq \frac{1}{2}$, modifying the result obtained for the standard oscillation scenario, where $\sin^2 \theta_{23} = \frac{1}{2}$. There are small differences observed for $n = 0$, which will be discussed later. From the bottom panel of Figure~\ref{fig:2d.dm32.s23.s13.dcp.c1.c2.c3.minos.plus.t2k}, which shows the $\delta_{\rm CP}-\sin^2\theta_{13}$ allowed regions, we note an effect on these regions due to some of the decoherence and relaxation scenarios, when compared to the standard oscillation model. The effect being smaller for $n = +2$ than for the other values of $n$. Such results are obviously dominated by the T2K $\nu_e+\bar{\nu}_e$ appearance signal which, as we know from Figure~\ref{spectrum.effect.of.gamma_0.minos.t2k}, is more sensitive to $n = -2$ and $n = 0$ than to $n = +2$.

\begin{figure}[ht]
\centering
\includegraphics[width=1.\textwidth]{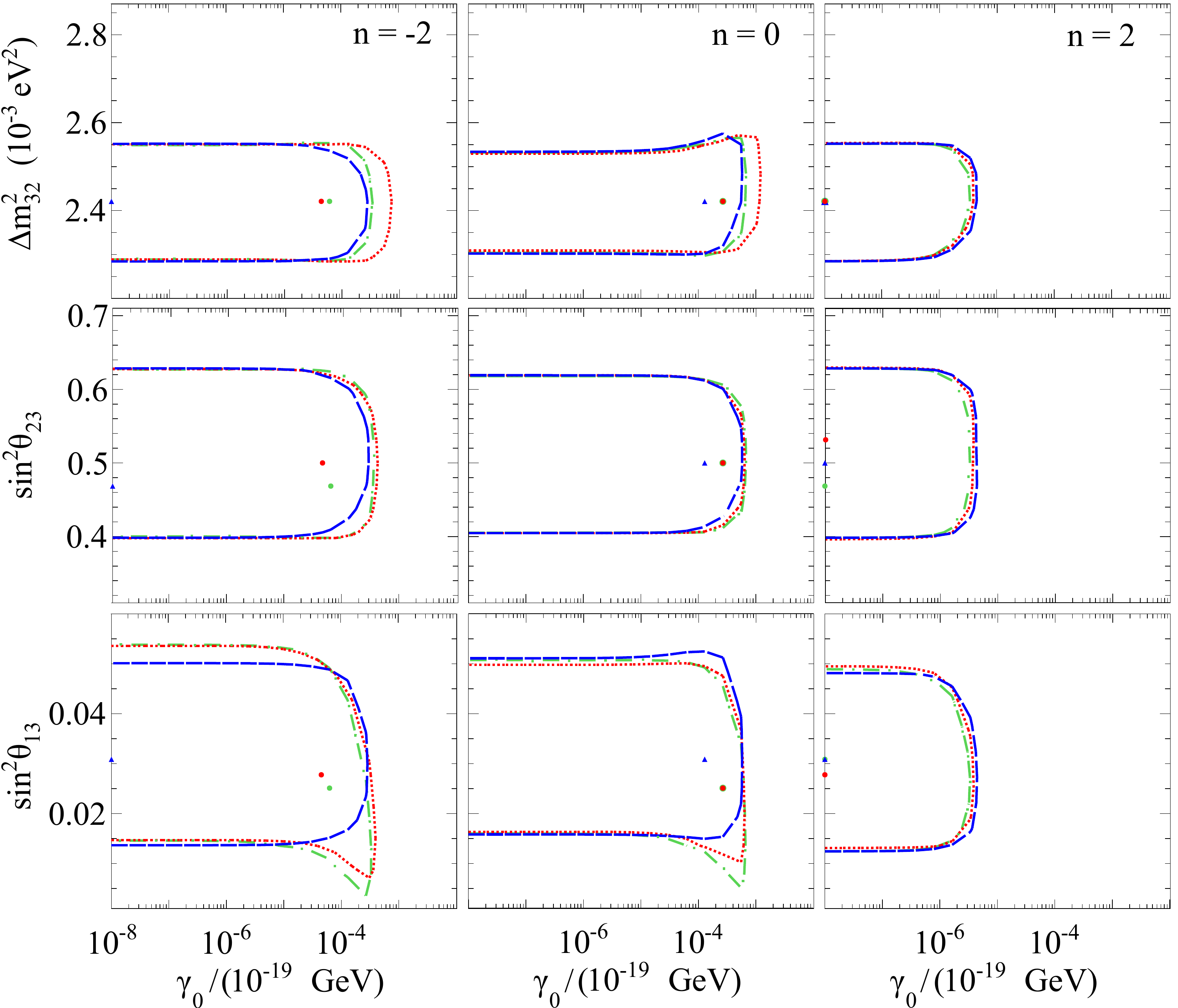}
\caption{The allowed regions of a given oscillation parameter and the $\gamma_0$ parameter, for 2 degrees of freedom. From the top to the bottom, we present the $\Delta m^2_{32}$, $\sin^2 \theta_{23}$, and $\sin^2 \theta_{13}$, respectively. The notation is the same as in Figure~\ref{fig:2d.dm32.s23.s13.dcp.c1.c2.c3.minos.plus.t2k}.}
\label{fig:2d.gamma_0.c1.c2.c3.minos.plus.t2k}
\end{figure}

In Figure~\ref{fig:2d.gamma_0.c1.c2.c3.minos.plus.t2k} we show the best-fit values and allowed regions, at the 90\% C.L., in the planes between an oscillation parameter and the $\gamma_0$ parameter, for the three cases and for the three values of $n$. These results contribute to better understand the effects on the contours presented in Figure~\ref{fig:2d.dm32.s23.s13.dcp.c1.c2.c3.minos.plus.t2k}. We show, in the upper and middle panels of Figure~\ref{fig:2d.gamma_0.c1.c2.c3.minos.plus.t2k}, the allowed regions of the planes $\Delta m_{32}^2$ -- $\gamma_{0}$ and $\sin^2 \theta_{23} - \gamma_0$, respectively. There are no significant modifications in the allowed regions among the cases, for each value of $n$, which gives confidence that these two oscillation parameters are robust with changes in the decoherence and relaxation scenario.
However, for $n = 0$ there is a small asymmetry on the $\Delta m_{32}^2$ component of the allowed region for values of $\gamma_0$ around the best-fit. This is related to the small distortion of the $\Delta m_{32}^2 - \sin^2 \theta_{32}$ allowed region, for $n = 0$, at Figure~\ref{fig:2d.dm32.s23.s13.dcp.c1.c2.c3.minos.plus.t2k}.

In the lower panel of Figure~\ref{fig:2d.gamma_0.c1.c2.c3.minos.plus.t2k}, we present the allowed region for the $\sin^2 \theta_{13} - \gamma_0$ plane at 90\% C.L. Due to some of the scenarios resulting in a best-fit value of $\gamma_0$ different from zero, as we have already discussed, there may be small distortions on the allowed region for the standard oscillation parameters, which is presented on Figure~\ref{fig:2d.dm32.s23.s13.dcp.c1.c2.c3.minos.plus.t2k}. That situation is, particularly, expressed on the $\sin^2\theta_{13}$ parameter for cases 1 and 2, with $n = -2$ and $0$, where the consequence is a decrease of the lower bound of $\theta_{13}$, for $\gamma_0$ values of a few of $10^{-23}$~GeV (Figure~\ref{fig:2d.gamma_0.c1.c2.c3.minos.plus.t2k}).

\subsection{Decoherence and relaxation bounds with and without the reactor constraint}

The $\gamma_0$ upper bounds for the combined MINOS and T2K analysis are presented at Table~\ref{table:summary-tableNO}, for each scenario. These bounds are dominated by the analysis of MINOS (T2K) data for $n=+2$ ($n=-2$). Once we combine the analysis of these two complementary experiments, with regard to our theoretical model, the resulting bounds are, naturally, less stringent than the best individual result. For instance, the result of MINOS for $n = +2$ is more stringent than the combined one, for every case.

The results for the combined analysis including the reactor constraint are also presented at Table~\ref{table:summary-tableNO}. There are no relevant differences in the bounds with and without the reactor constraint for each scenario. However, we notice that the differences for $n = -2$ and $0$ are larger than for $n = +2$, due to the effect on $\theta_{13}$ previously discussed. Obviously, the reactor constraint affects $\theta_{13}$, causing a stronger effect on the scenarios better constrained by T2K 
$\nu_e + \overline{\nu}_e$ appearance data.

\begin{figure}[ht]
\centering
\includegraphics[width=1.\textwidth]{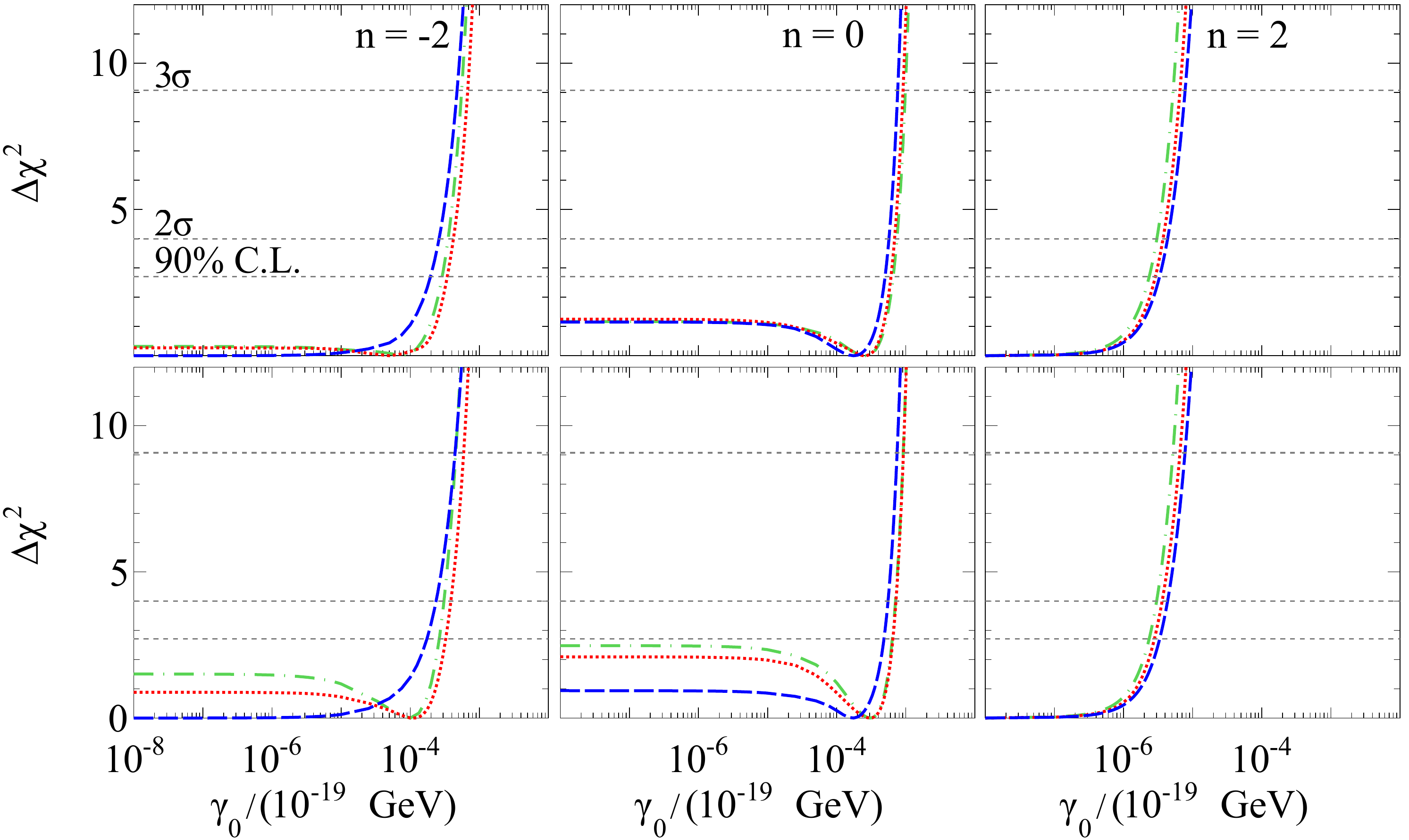} \hfill
\vspace{-0.8cm}
\caption{The projections of $\Delta\chi^2$ as a function of $\gamma_0$ for the combined analysis of MINOS and T2K dataset without (upper panel) and with (lower panel) the reactor constraint. The notation is the same as in  Figure~\ref{fig:2d.dm32.s23.s13.dcp.c1.c2.c3.minos.plus.t2k}.}
\label{fig:1d.gamma_0.c1.c2.c3.minos.comb.t2k}
\end{figure}

The upper (lower) panel of Figure~\ref{fig:1d.gamma_0.c1.c2.c3.minos.comb.t2k} shows the projection of $\Delta\chi^2$ as a function of $\gamma_0$ for each scenario and value of $n$, considering the combined analysis without (with) the reactor constraint. For $n = -2$, cases 1 and 2 show a slight preference for the $\gamma_0$ parameter to be non-zero as best-fit value. For $n = 0$, all three cases show also a preference for $\gamma_0$ value different from zero. The significance of those non-zero best-fit values is increased by the inclusion of the constraint from the reactor data analysis. The results for $n = +2$, which is dominated by the analysis of the MINOS data, show no visible effect due to the reactor constraint. All the bounds at the 90\% C.L. for the combined analyses presented at Table~\ref{table:summary-tableNO} were obtained from these plots.

\subsection{Comparing our results with previous bounds}

In this section we compare our results with some previous bounds on $\gamma_0$ from the literature, which are presented at Table~\ref{table:summary-tableNO}. These bounds are coming from phenomenological or sensitivity analyses of data from Super-Kamiokande~\cite{Lisi:2000zt}, MINOS~\cite{deOliveira:2013dia}, KamLAND~\cite{Gomes:2016ixi}, IceCube/DeepCore~\cite{Coloma:2018idr}, and DUNE~\cite{Gomes:2018inp} (under two different configurations). All these results considered the normal mass hierarchy, but were based on different confidence levels, which allow comparisons in terms of orders of magnitude only.

Our results for $n = -2$ are the best constraints for $\gamma_0$ in the literature, for any of the cases analysed, $\gamma_0 \sim 10^{-23}$ GeV. We obtained limits one order of magnitude better than the previous best bound~\cite{deOliveira:2013dia}, for MINOS, and three (five) orders of magnitude better than for DeepCore (IceCube)~\cite{Coloma:2018idr}.

For the energy independent scenarios ($n = 0$), our bounds on $\gamma_0$ ($\sim 10^{-23}$~GeV) are better than or similar to the bounds from the data of other experiments, except from the IceCube data, which is the best limit by one order of magnitude. Hence, our result does not exclude the inferred value of Ref.~\cite{Coelho:2017zes}, which claims that a decoherence of strength $(2.3 \pm 1.1) \times 10^{-23}$~GeV could solve a previous tension on $\theta_{23}$ measurements between NOvA and T2K. However, that value was already excluded by the limits from IceCube data and could be excluded by DUNE, accordingly to the sensitivity analysis from Ref.~\cite{Gomes:2018inp}. It is expected that the sensitivity for the high energy flux configuration of DUNE would result in the best limit on $\gamma_0$ for energy independent decoherence, by one order of magnitude better than the IceCube limit.

We also point out that, for $n = 0$, there is an interesting tension between the IceCube and our results. The non-zero best-fit values of $\gamma_0$ we obtained, with significance ranging from 68\% to 90\%~C.L. (Figure~\ref{fig:1d.gamma_0.c1.c2.c3.minos.comb.t2k}) are excluded by the IceCube limit (at 95\%~C.L.). This conflict could be clarified by another analysis, for instance, by the future DUNE experiment.

Concerning the results for $n = +2$, our bounds are comparable to the previous bounds from the analyses of MINOS and DeepCore data. The best limits, however, are from the analyses of Super-Kamiokande and IceCube data, which are around three and six orders of magnitude, respectively, more stringent than our bounds.

\section{Conclusions}
\label{sec:conclusions}

We have performed a phenomenological analysis and presented limits to neutrino quantum decoherence and relaxation for a range of possible scenarios, using the MINOS and T2K long-baseline data. The formalism of an open quantum system was applied to neutrinos and anti-neutrinos on the survival and transition probabilities. The study of the oscillatory and non-oscillatory terms of the probability allows the investigation of the effect of both decoherence and relaxation.

Three scenarios were investigated. In the first one, all decoherence parameters are equal and we allow the possibility of relaxation. The second one is the same as the first, but no relaxation is allowed. And in the third one, we consider only the decoherence parameters related to the atmospheric sector and no relaxation is allowed. We assume an energy dependence of the decoherence parameter to be parameterized as $\Gamma=\gamma_0 (E/\mbox{GeV})^n$, with $n = -2, 0,$ and $+2$. Obviously, the models with an energy dependence on intermediate values of $n$, such as $\pm 1$, are contained on the ranges presented for each scenario.

The complementary behaviour of MINOS and T2K with regard to our theoretical framework was clear in the analyses we performed. The individual analysis of MINOS (T2K) data resulting in more stringent bounds on $\gamma_0$ for $n = +2$ ($n = -2$) than for the other values of $n$.

We have found that the decoherence and relaxation scenarios result in small distortions on the allowed regions of the oscillation parameters. The more relevant impact is on $\sin^2 \theta_{13}$, due to the effect of $\gamma_0$ in the T2K $\nu_e + \overline{\nu}_e$ appearance analyses. For some of the scenarios we obtained non-zero best-fit values of $\gamma_0$, which contribute to the observed effect on the oscillation parameters. The inclusion of a reactor constraint on $\theta_{13}$ has a small impact on our results.

In both individual and combined analyses, we clearly observe that, for each value of $n$, there are no significant differences among the bounds on $\gamma_0$ for the three decoherence and relaxation scenarios investigated. Thus, we conclude that the data we analysed are not sensitive to: (i) the effect of relaxation, when comparing scenarios 1 and 2; and (ii) the effect of constraining or not the decoherence parameters between the solar and atmospheric sectors, when comparing scenarios 2 and 3. In other words, {\it the results are independent of the scenarios we investigate}.

{\it Concerning the bounds on $\gamma_0$, our analysis presents the best limits in the literature for the energy dependence with $n = -2$}. The upper bound from the combined analysis, including the reactor constraint, for the scenario 3, is $\gamma_0 < 1.7 \times 10^{-23}$~GeV, at the 90\%~C.L., which improves the previous best limit in one order of magnitude. Our results for $n = 0$ and $+2$ are similar to the other bounds for long-baseline data. For those values of $n$, the best bounds on $\gamma_0$ are from atmospheric data analyses.

It is worth mentioning that, for some scenarios, the non-zero best-fit values of $\gamma_0$ we obtained, with significance ranging from 68\% to 90\% confidence levels, are excluded by the IceCube limits (at 95\% C.L.). For instance, for the energy independent scenarios $(n = 0)$, our best-fit values of $\gamma_0$, which are consistent to the value considered on Ref.~\cite{Coelho:2017zes} to explain a previous tension between NOvA and T2K, are excluded by the IceCube bounds on $\gamma_0$.
The tension on those results claims for a new analysis that could potentially clarify the conflict.

\begin{acknowledgments}
The authors thanks Carlos Arguelles for valuable discussion about the paper. R.A.G.~was supported by FAPEG and by CNPq grants 307334/2019-8 and 310708/2022-2. O.L.G.P. was supported by FAPESP funding Grant 2016/08308-2, FAEPEX funding grant 2391/2017 and 2541/2019, and CNPq grants 306565/2019-6 and 306405/2022-9. R.A.G. and O.L.G.P. are thankful for the support of FAPESP funding Grant 2014/19164-6. This study was funded in part by the Coordena\c{c}\~ao de Aperfei\c{c}oamento de Pessoal de N\'ivel Superior - Brasil (CAPES) - Finance Code 001.
\end{acknowledgments}

\appendix
\section{Some properties of the decoherence and relaxation neutrino system}
\label{deco-prop}

The Gorini-Kossakowski-Sudarshan-Lindblad  equation~\cite{Gorini:1975nb,Lindblad1976} is a very general equation for systems interacting with a larger system, so-called the environment. We will assume some general conditions:
\begin{enumerate}
\item The Von Neumann entropy of the subsystem is always positive, which implies that the operators  $V_{\epsilon}$ are hermitian, $V_{\epsilon}=V_{\epsilon}^{\dagger}$~\cite{Benatti1988,Phdoliveira}, 
or that $\sum_{\epsilon} V_{\epsilon} V_{\epsilon}^{\dagger} = I$~\cite{Oliveira:2013nua}. 
With this condition  we can use the following expansions:
\begin{eqnarray}
\mathcal{H}=\sum_{\mu} \mathcal{H}_{\mu} F_{\mu},\quad V_{\epsilon}  =  \sum_\mu v^{(\epsilon)}_{\mu} F_{\mu}, \quad  \rho  =  \sum_\mu \rho_{\mu} F_{\mu}, 
\end{eqnarray}
where the $F_{\mu}$ matrices are $F_0 = \frac{1}{\sqrt{6}}I_3$ and $F_j = \frac{1}{2}\lambda_j$, where $\lambda_j$ are the Gell-Mann matrices and $j=(1,\cdots,8)$. The dissipative term can be written as  
\begin{eqnarray}
\mathcal{D}[\rho(t)]= \sum_{\alpha\beta} D_{\alpha\beta} \rho_{\beta}F_{\alpha}, \quad \mathcal{D}_{\alpha\beta}\equiv\frac{1}{2}\sum_{\mu\nu\gamma} \left(\vec{v}_{\mu}\cdot\vec{v}_{\nu}\right)f_{\gamma\alpha\mu}f_{\gamma\nu\beta}, 
\label{dexplicit}
\end{eqnarray}
where $\vec{v}_{\mu}\cdot\vec{v}_{\nu}\equiv \sum _{\epsilon} v_{\mu}^{(\epsilon)}v_{\nu}^{(\epsilon)} $, and $f_{\alpha\mu\gamma}$ is equal to zero, for $\alpha,\mu,\gamma =0$ and equal to SU(3) structure constants, for 
$\alpha,\mu,\gamma=1,2,3$, coming from the following relation
\begin{eqnarray}
[F_i, F_j] = i\sum_k f_{ijk}F_k, \quad 
\end{eqnarray}
with $i,j,k=(1,\cdots,8)$.
\item Probability conservation: 
We will impose probability conservation, following Ref.~\cite{Phdoliveira},
\begin{eqnarray}
{\rm Tr}\left(\rho(t) \right)=1 \longrightarrow \mathcal{D}_{\mu 0}=\mathcal{D}_{0\mu}=0,\quad  
\label{probcons}
\end{eqnarray}
with $\mu = 1, 2, 3$. Under these conditions the Gorini-Kossakowski-Sudarshan-Lindblad equation, defined in Eq.~(\ref{eq:lindblad.equation}), can be rewritten using the Eq.~(\ref{probcons}) and (\ref{dexplicit}) as
\begin{eqnarray}
\dot{\rho}_{i} = \sum_{k,j} \left( f_{ikj}\mathcal{H}_k + \mathcal{D}_{ij} \right) \rho_{j}, \quad  \rho_0=\sqrt{2/3},\quad 
\label{eq:lindblad.expanded.su3}
\end{eqnarray}
where an explicitly symmetric form for the $\mathcal{D}_{ij}$ is given by 
\begin{eqnarray}
\mathcal{D}_{ij} =\frac{1}{2}\left[\sum_{p,l,q=1}^{8} 
\left(\vec{v}_{p} \cdot \vec{v}_{l} \right) f_{qip}f_{qlj}
 \right]
 =-\frac{1}{4}\left[\delta_{ij} \sum_{p=1}^8 \vec{v}_p \cdot \vec{v}_p - \frac{1}{3} \vec{v}_i \cdot \vec{v}_j  \right]. 
\label{eq:matriz.dissipacao3}
\end{eqnarray}
We have used the property of the SU(3) structure constants $f_{ikj}=i[T_{i}]_{kj}$, in which $T_{i}$ is the adjoint representation of SU(3) algebra. Using the properties of products of Gell-Mann matrices, we obtain 
$\sum_q f_{qip}f_{qlj}=(-1) \left( T_q\right)_{ip}\left(T_q\right)_{lj}=\frac{1}{2} \left[\delta_{ij}\delta_{pl}-\frac{1}{3} \delta_{ip}\delta_{lj}\right]$, where the details are given in Ref.~\cite{Phdsandro}. For the assumed form of the $D$ matrix in Eq.~(\ref{eq:matrix.M}), see Ref.~\cite{Phdabner}, the probability conservation implies that $D_{ii}<0$ for all $i$.

\item Complete positivity: 
In general, the evolution given by Eq.~(\ref{eq:lindblad.expanded.su3}) will have a formal solution~\cite{Gago:2002na} as
\begin{eqnarray}
\rho(t)=T e^{ \mathcal{M}_{\rm diag}^{\prime} t} T^{-1} \rho(t=0),
\end{eqnarray}
where $\mathcal{M}_{\rm diag}^{\prime}\equiv (
\lambda_1,\lambda_2,\cdots,\lambda_8)$ is the diagonal form of the $\mathcal{M}$, defined in Eq.~(\ref{eq:matrix.elements.M}), $T$ matrix are the eigenvectors and $\lambda_i$ are the eigenvalues of $\mathcal{M}$. In case the eigenvalues of $\mathcal{M}$ are positive, then the probability would have exponential growth behavior, that would violate the probability unitarity. The requirement to have only physically viable solutions, with negative eigenvalues, is called {\emph complete positivity}~\cite{Gorini:1975nb,Lindblad1976}.  A detailed discussion on the implications of the complete positivity is given by Ref.~\cite{Phdsandro}.

\item Condition for energy exchange conservation:
The solutions of Eq.~(\ref{eq:lindblad.expanded.su3}) can be classified in two classes:

\begin{enumerate}
\item no energy exchange between the system and the environment. This statement can be written as  $[\mathcal{H},V_k]=0$, and was adopted, for example, in Refs.~\cite{Benatti:2000ph,Gago:2000qc,Gago:2000nv,Lisi:2000zt,Benatti:2001fa,Gago:2002na,Phdemelo,DeMelo:2003yg,Morgan:2004vv,Hooper:2004xr,Barenboim:2006xt,Fogli:2007tx,Farzan:2008zv,Oliveira:2010zzd,deOliveira:2013dia,Oliveira:2013nua,Oliveira:2014jsa,Bakhti:2015dca,Oliveira:2016asf,Gomes:2016ixi,Coelho:2017zes,Coelho:2017byq,Carpio:2017nui,Mosquera:2017vir,Capolupo:2018hrp,Coloma:2018idr,Gomes:2018inp,Mosquera:2018yar,Carpio:2018gum,Carrasco:2018sca,BalieiroGomes:2018koe,deHolanda:2019tuf,Carrasco-Martinez:2020mlg,deGouvea:2021uvg,FigueiredoSeverianoAlves:2020jue,Buoninfante:2020iyr}.  It is the case where we have only decoherence. The Hamiltonian $\mathcal{H}$ in the mass basis can be written as 
\begin{eqnarray}
\mathcal{H} & = &\mathcal{H}_0 F_0+ \mathcal{H}_3 F_3+ \mathcal{H}_8 F_8,  \label{hmatrix1}
\end{eqnarray}
where
\begin{eqnarray}
\mathcal{H}_3&=&-\Delta_{21}, \mbox{ and ~}
\mathcal{H}_8=\frac{\Delta_{21}-2\Delta_{31}}{2\sqrt{3}}.
\label{hmatrix2}
\end{eqnarray}
The $\mathcal{H}_0$ term is a constant matrix, not relevant for us, and $\Delta_{ij}\equiv \Delta m^2_{ij}/2E$, with $\Delta m^2_{ij}\equiv m^2_i-m^2_j$. The condition to have no energy exchange is that the expected value of Hamiltonian to be time independent, which implies that $\sum_i \mathcal{H}_i \mathcal{D}_{ii}=0 \to \mathcal{D}_{33}=\mathcal{D}_{88}=0$. 
\item energy exchange is possible and then $\mathcal{D}_{33},~\mathcal{D}_{88}\neq  0$. This approach was used in Ref.~\cite{Benatti:2000ph,Barenboim:2004wu,Abbasi:2009nfa,Phdoliveira,Oliveira:2014jsa,Oliveira:2016asf,Phdsandro,Richter-Laskowska:2018ikv,Phdabner,Buoninfante:2020iyr}.
\end{enumerate}
\end{enumerate}

\subsection{Probability Computation}
\label{proba-compute}

The computation of the probability is very well described in Ref.~\cite{Gago:2002na}. The Ref.~\cite{Richter-Laskowska:2018ikv} made available a Mathematica code to compute numerically the decoherence probability for different cases. In our case we solve analytically the Eq.~(\ref{eq:lindblad.expanded.su3}) using the explicit form of $\mathcal{M}$ given in Eq.~(\ref{eq:matrix.M}). 
In the flavor basis, the initial condition at $t = 0$, for a flavor state $\alpha$, can be described as 
\begin{eqnarray}
\rho^{\alpha}(t=0) &\equiv  &\sum_{ij}U^{*}_{\alpha i}U_{\alpha j}|\nu_i\rangle \langle \nu_j|=\left(
\begin{array}{ccc}
|U_{\alpha 1}|^2 & U_{\alpha 1}^*U_{\alpha 2} & U_{\alpha 1}^* U_{\alpha 3} \\
U_{\alpha 2}^*U_{\alpha 1}  & |U_{\alpha 2}|^2 & U_{\alpha 2}^* U_{\alpha 3}  \\
U_{\alpha 3}^*U_{\alpha 1}  & U_{\alpha 3}^*U_{\alpha 2} &    |U_{\alpha 3}|^2\\
\end{array}
\right) \nonumber \\
&=& \left( \begin{array}{ccc}
	\frac{1}{\sqrt{3}} + \frac{1}{2}(\rho^{\alpha}_{3} + \frac{1}{\sqrt{3}}\rho^{\alpha}_{8}) & \frac{1}{2}(\rho^{\alpha}_{1} -i\rho^{\alpha}_{2}) & \frac{1}{2}(\rho^{\alpha}_{4} -i\rho^{\alpha}_{5}) \\
	\frac{1}{2}(\rho^{\alpha}_{1} + i\rho^{\alpha}_{2}) & \frac{1}{\sqrt{6}}\rho^{\alpha}_{0} - \frac{1}{2}(\rho^{\alpha}_{3} - \frac{1}{\sqrt{3}}\rho^{\alpha}_{8}) & \frac{1}{2}(\rho^{\alpha}_{6} - i\rho^{\alpha}_{7}) \\
	\frac{1}{2}(\rho^{\alpha}_{4} + i\rho^{\alpha}_{5}) & \frac{1}{2}(\rho^{\alpha}_{6} + i\rho^{\alpha}_{7}) & \frac{1}{\sqrt{3}}-\frac{1}{\sqrt{3}}\rho^{\alpha}_{8} \end{array} \right)_{t=0}, \nonumber \\
\label{udoeh}
\end{eqnarray}
where $U$ is the PMNS neutrino mixing matrix~\cite{1962PThPh..28..870M,Pontecorvo:1957cp} and 
$\rho^{\alpha}_i(t=0)$, with $i=1,\cdots,8,$ are the components of density matrix in SU(3) basis.

With these initial conditions the solution for the components $\rho^{\alpha}_i(L)$, from Eq.~(\ref{eq:lindblad.expanded.su3}), for the given form of $\mathcal{M}$ is as follows
\begin{eqnarray}
\rho^{\alpha}_{1,2}(L)&=&e^{-\Gamma_{21}L}
\left\{ \rho^{\alpha}_{1,2}(0) \left(\mp \frac{\Delta\mathcal{D}_{21}}{\Omega_{21}}\sin \frac{\Omega_{21}L}{2}+
\cos \frac{\Omega_{21}L}{2}\right)
\mp \rho^{\alpha}_{2,1}(0) \left(\frac{2\Delta_{21}}{\Omega_{21}}\right)\sin \frac{\Omega_{21}L}{2}\right\}, \nonumber \\
\rho^{\alpha}_{4,5}(L)&=&e^{-\Gamma_{31}L}
\left\{ \rho^{\alpha}_{4,5}(0) \left(\mp \frac{\Delta\mathcal{D}_{31}}{\Omega_{31}}\sin \frac{\Omega_{31}L}{2}+
\cos \frac{\Omega_{31}L}{2}\right)
\mp \rho^{\alpha}_{5,4}(0) \left(\frac{2\Delta_{31}}{\Omega_{31}}\right)\sin \frac{\Omega_{31}L}{2}\right\}, \nonumber \\
\rho^{\alpha}_{6,7}(L)&=&e^{-\Gamma_{32}L}
\left\{ \rho^{\alpha}_{6,7}(0) \left(\mp \frac{\Delta\mathcal{D}_{32}}{\Omega_{32}}\sin \frac{\Omega_{32}L}{2}+
\cos \frac{\Omega_{32}L}{2}\right)
 \mp \rho^{\alpha}_{7,6}(0) \left( \frac{2\Delta_{32}}{\Omega_{32}}\right)\sin \frac{\Omega_{32}L}{2}\right\}, \nonumber \\
\rho^{\alpha}_{8}(L) &=&
e^{\mathcal{D}_{88}L}
\rho^{\alpha}_{8}(0),\quad \rho^{\alpha}_{3}(L)=
e^{\mathcal{D}_{33}L}
\rho^{\alpha}_{3}(0),
 \label{final:eqq}
\end{eqnarray}
where we define the $\Gamma_{ij}$, 
\begin{eqnarray}
\Gamma_{21} = - \left(\frac{\mathcal{D}_{11} + \mathcal{D}_{22}}{2}\right), \quad
\Gamma_{31} = - \left(\frac{\mathcal{D}_{44} + \mathcal{D}_{55}}{2}\right), \quad
\Gamma_{32} = - \left(\frac{\mathcal{D}_{66} + \mathcal{D}_{77}}{2}\right),
 \label{eq:gamma.sector3}
 \end{eqnarray}
 and the combination
 \begin{eqnarray}
 \Omega_{21} = \sqrt{4\Delta^2_{21} - (\Delta\mathcal{D}_{21})^2}, \hspace{2.0mm}
 \Omega_{31} = \sqrt{4\Delta^2_{31} - (\Delta\mathcal{D}_{31})^2}, \hspace{2.0mm}
 \Omega_{32} = \sqrt{4\Delta^2_{32} - (\Delta\mathcal{D}_{32})^2}, \hspace{2.5mm}
 \label{eq:gamma.sector1} 
 \end{eqnarray}
 where $\Delta\mathcal{D}_{ij}$ is 
\begin{eqnarray}
 \Delta\mathcal{D}_{21} = \mathcal{D}_{22} -\mathcal{D}_{11}, 
 \quad 
 \Delta\mathcal{D}_{31} = \mathcal{D}_{55} -\mathcal{D}_{44},
 \quad 
 \Delta\mathcal{D}_{32} = \mathcal{D}_{77} -\mathcal{D}_{66}.
 \label{eq:gamma.sector2}
\end{eqnarray}

The probability now can be computed as 
\begin{eqnarray}
P(\nu_{\alpha} \to \nu_{\beta})\equiv{\rm Tr}\left( 
\rho^{\alpha} (t=0) \rho^{\beta}(t)\right)=\sum_{i=0}^{8} \rho^{\beta}_i(t)\rho^{\alpha}_i(t=0),
 \label{eq:tudo} 
\end{eqnarray} 
where, in the last equality, we should put the explicit expression for $\rho_i(t)$ from Eq.~(\ref{final:eqq}) and the initial conditions from Eq.~(\ref{udoeh}). We then get the full probability as
\begin{eqnarray}
P(\nu_{\alpha} \to \nu_{\beta}) & = &
 \delta_{\alpha\beta}   +2\sum_{j>i}\bigg{ \{ }\mathbb{R}[{\rm W}_{\alpha\beta}^{ij}] 
\left[\cos\left( \frac{\Omega_{ij}}{2}L\right)-1\right] \nonumber\\
&+& \left[\frac{\mathbb{R}[{\rm Y}_{\alpha\beta}^{ij}](\Delta\mathcal{D})_{ij}  
- \Im [{\rm W}_{\alpha\beta}^{ij}]2\Delta_{ji}}{\Omega_{ij}} \right]\sin\left( \frac{\Omega_{ij}}{2}L \right)\bigg{\}}e^{-\Gamma_{ij}L} \nonumber\\
&-&\frac{1}{6}\left( 1 - 3|U_{\alpha 3}|^2 \right)\left( 1 - 3|U_{\beta 3}|^2 \right)\left(1-e^{\mathcal{D}_{88}L}\right)  \nonumber\\
& - & \frac{1}{2}\left( |U_{\alpha 1}|^2 - |U_{\alpha 2}|^2 \right)\left( |U_{\beta 1}|^2 - |U_{\beta 2}|^2 \right) \left(1- e^{\mathcal{D}_{33}L}\right).
 \label{eq:prob.transicao.3.sabores.osc.dec_termos_U_apendice}
\end{eqnarray}
An expression for the decoherence probability to be readable when compared to the usual three neutrino probability is shown in Eq.~(\ref{eq:prob.transicao.3.sabores.osc.dec_termos_U}).

\section{Expected Events}
\label{app:validation}

We have calculated the expected number of events $N^{\rm mod}$ for an energy bin $i$ and for a certain theoretical model, as follows:
\begin{eqnarray}
N^{\rm mod}_i = \left(\sum^{\rm bins}_{\alpha=1} \phi^{\rm Far}_\alpha \times \overline{P^{\rm mod}_\alpha} \times \sigma^{\rm int}_\alpha \times G_{\alpha i}\right) \times \epsilon_i
\label{eq:calculated.events}.
\end{eqnarray}
We perform a sum over all bins $\alpha$ to consider their contribution to a specific bin $i$ due to the smearing matrix $G$ used to transform the true energy $E_\alpha$ to the reconstructed energy $E_i$, as described below. Here $\phi^{\rm Far}$ is the neutrino flux at the Far detector, which we have calculated as described at Ref.~\cite{Gago:2017zzy}. $\overline{P}^{\rm mod}_{\alpha}$ is the average probability per bin for the model being investigated, obtained by
\begin{eqnarray}
\overline{P^{\,\rm mod}_{\alpha}} = \frac{1}{\delta_\alpha}\int^{E_\alpha + \delta_\alpha/2}_{E_\alpha - \delta_\alpha/2} P^{\rm mod}(E) dE,
\end{eqnarray}
where $E_{\alpha}$ and $\delta_{\alpha}$ are the central energy and width of the bin, respectively, and $P^{\rm mod}(E)$ is the probability formula as a function of the true energy of the neutrino. The cross-section for a certain interaction is given by $\sigma^{\rm int}_\alpha$, and the detection efficiency is described by $\epsilon$, which is a function of the reconstructed energy.

The $G_{\alpha i}$ are the elements of the transformation matrix, modeled by Gaussian functions as follows:
\begin{eqnarray}
G_{\alpha i} = \frac{1}{N} \text{exp}\left[ -\frac{1}{2}\left(\frac{E_i - E_\alpha + \delta E}{\sigma^{G}_\alpha}\right)^2\right]\label{eq:reconstrution. energy. matrix},
\end{eqnarray}
where $N$ is a normalization constant, $E_i$ is the reconstructed energy, and $E_\alpha$ is the true energy. For \textit{non-quasi elastic} processes we consider a shift $\delta E$ in the Gaussian function to handle the problem to determine the neutrino energy. To obtain the smearing matrix we used two Gaussian functions to model it in an asymmetric shape. The Gaussian resolution $\sigma^G_{\alpha}$ used in our analysis is described at Ref.~\cite{Gago:2017zzy}, except for the T2K $\nu_e$ appearance data, in which we used the following resolutions for the neutrino run, 
\begin{eqnarray}
 \sigma_{\alpha}^{\nu_e, r} &=& 1.97 E^2_\alpha - 1.98 E_\alpha + 0.53~{\rm (GeV)},\nonumber\\
 \sigma_{\alpha}^{\nu_e, l} &=& 0.13 E_\alpha~{\rm (GeV)}\label{eq:resolution.nue.neutrino-run.t2k},
\end{eqnarray}
and for the anti-neutrino run,
\begin{eqnarray}
 \sigma_{\alpha}^{\nu_e, r} &=& 2.33 E^2_\alpha - 2.17 E_\alpha + 0.43~{\rm (GeV)},\nonumber\\
 \sigma_{\alpha}^{\nu_e, l} &=& 0.10 E_\alpha~{\rm (GeV)}
 \label{eq:resolution.nue.antineutrino-run.t2k},
\end{eqnarray}
where the index \textit{l}(\textit{r}) represents the resolution of the matrix which smear the events from higher (lower) to lower (higher) energies. The validation of this method under the standard oscillation model for the dataset used in this analysis is presented in Ref.~\cite{Gomes:2014yua, Gago:2017zzy}.

\bibliographystyle{JHEP}
\bibliography{decoherence_minos_t2k_paper}

\end{document}